\documentclass{PoS-hep}
\usepackage{amsmath}
\usepackage{bm}
\usepackage{epsfig}
\usepackage{graphics}
\usepackage{upgreek}
\usepackage{amsfonts}
\usepackage{amsbsy}
\usepackage{amscd}
\usepackage{bbm}
\usepackage{eufrak}
\usepackage{cite}

\renewenvironment{subequations}{%
\refstepcounter{equation}%
\setcounter{parentequation}{\value{equation}}%
  \setcounter{equation}{0}
  \ignorespaces
}{%
  \setcounter{equation}{\value{parentequation}}%
  \ignorespacesafterend
}

\newcommand{\beqs}{\begin{subequations}}
\newcommand{\eeqs}{\end{subequations}}
\newcommand{\eec}{\end{center}}
\newcommand{\bec}{\begin{center}}
\newcommand{\eem}{\end{matrix}}
\newcommand{\bem}{\begin{matrix}}
\newcommand{\Eref}[1]{Eq.~(\ref{#1})}
\newcommand{\Sref}[1]{Sec.~\ref{#1}}
\newcommand{\Fref}[1]{Fig.~\ref{#1}}
\newcommand{\Tref}[1]{Table~\ref{#1}}
\newcommand{\cref}[1]{Ref.~\cite{#1}}

\newcommand\eqs[2]{Eqs.~(\ref{#1}) and (\ref{#2})}

\newcommand\eqss[3]{Eqs.~(\ref{#1}), (\ref{#2}) and (\ref{#3})}
\newcommand{\sEref}[2]{Eq.~(\ref{#1}{\ftn\sf {#2}})}
\newcommand{\sFref}[2]{Fig.~\ref{#1}-{\ftn\sf ({#2})}}

\newcommand{\ck}{\ensuremath{c_R}}
\newcommand{\Fcr}{\ensuremath{F_{R}}}
\newcommand{\fr}{\ensuremath{f_{R}}}
\newcommand{\fk}{\ensuremath{f_{K}}}
\newcommand{\fkk}{\ensuremath{f_K}}
\newcommand{\fp}{\ensuremath{f_\Phi}}
\newcommand{\fsp}{\ensuremath{f_{S\Phi}}}
\newcommand{\Fk}{\ensuremath{F_K}}
\newcommand{\ks}{\ensuremath{k_S}}
\newcommand{\ksp}{\ensuremath{k_{S\Phi}}}
\newcommand{\kpp}{\ensuremath{k_{\Phi}}}

\newcommand{\kspb}{\ensuremath{\bar k_{S\Phi}}}
\newcommand{\kppb}{\ensuremath{\bar k_{\Phi}}}
\newcommand{\kppp}{\ensuremath{k_{\Phi\bar\Phi}}}

\newcommand{\eeq}{\end{equation}}
\newcommand{\beq}{\begin{equation}}
\newcommand{\ba}{\begin{array}}
\newcommand{\ea}{\end{array}}
\newcommand{\bea}{\begin{eqnarray}}
\newcommand{\eea}{\end{eqnarray}}
\newcommand{\ftn}{\footnotesize}

\newcommand{\etal}{{\it et al.\/}}

\newcommand{\astroph}[1]{{\ftn \tt astro-ph/#1}}
\newcommand{\arxiv}[1]{{\ftn\tt  arXiv:#1}}
\newcommand{\ns}{\ensuremath{n_{\rm s}}}

\newcommand\vev[1]{\langle {#1} \rangle}
\def\lf{\left(}
\def\rg{\right)}

\newcommand{\Vhi}{\ensuremath{\widehat{V}_{\rm I}}}
\newcommand{\Hhi}{\ensuremath{\widehat H_{\rm I}}}
\newcommand{\as}{\ensuremath{\alpha_{\rm s}}}

\newcommand{\rcc}{\ensuremath{\emph{R}}}
\newcommand{\rce}{\ensuremath{\widehat{{R}}}}
\newcommand{\Ve}{\ensuremath{\widehat{V}}}
\newcommand{\Ne}{\ensuremath{\widehat{N}}}

\def\aal{{\bar\alpha}}
\def\bbet{{\bar\beta}}
\def\al{{\alpha}}
\def\bt{{\beta}}

\def\n{\bar{n}}

\newcommand{\Trh}{\ensuremath{T_{\rm rh}}}

\newcommand{\ld}{\ensuremath{\lambda}}

\newcommand{\Ldf}{\ensuremath{\Lambda_{\rm eff}}}

\newcommand{\se}{\ensuremath{\widehat\phi}}
\newcommand{\sex}{\ensuremath{\widehat{\phi}_*}}

\newcommand{\geu}{\ensuremath{\widehat g}}

\newcommand{\mP}{\ensuremath{m_{\rm P}}}

\newcommand{\sg}{\ensuremath{\phi}}
\newcommand{\sig}{\ensuremath{\phi}}
\renewcommand{\sigma}{\ensuremath{\phi}}
\newcommand{\sigf}{\ensuremath{\phi_{\rm f}}}

\newcommand{\xsg}{\ensuremath{x_{\phi}}}

\newcommand{\GeV}{{\mbox{\rm GeV}}}
\newcommand{\what}{\ensuremath{\widehat}}
\newcommand{\Khi}{\ensuremath{K}}

\newcommand{\Vhio}{\ensuremath{\widehat V_{\rm I0}}}
\newcommand{\Ohi}{\ensuremath{\Omega}}
\newcommand{\He}{\ensuremath{\widehat{H}}}
\newcommand{\diag}{\mbox{\sf\ftn diag}}

\newcommand{\Mgut}{\ensuremath{M_{\rm GUT}}}
\newcommand{\Ggut}{\ensuremath{G_{\rm GUT}}}
\newcommand{\Gsm}{\ensuremath{G_{\rm SM}}}
\newcommand{\Mpq}{\ensuremath{M}}
\newcommand{\mpq}{\ensuremath{m_{BL}}}
\newcommand{\ldu}{\ensuremath{\uplambda}}

\def\th{{\theta}}
\def\thb{{\bar\theta}}
\def\thn{{\theta_{\Phi}}}

\title{\Large Models of Non-Minimal Chaotic Inflation in Supergravity}

\ShortTitle{Models of nMI in SUGRA}

\author{\speaker{C. Pallis}\\
        Department of Physics, University of Cyprus, \\
        P.O. Box 20537, Nicosia 1678, CYPRUS\\
        E-mail: \email{cpallis@ucy.ac.cy}}

\abstract{We show how we can implement chaotic inflation in the
context of supergravity by conveniently selecting the functional
form of a strong enough non-minimal coupling between the inflaton
and the Ricci scalar curvature. The procedure can be applied when
a gauge singlet or non-singlet inflaton is coupled to another
singlet superfield within linear-quadratic, trilinear or bilinear
superpotential terms. The tachyonic instability occurring along
the direction of the accompanying non-inflaton field can be cured
by expanding the kinetic part of the frame function up to the
fourth order in powers of the various fields. In the case of a
gauge non-singlet inflaton, though, a conjugation symmetry has to
be imposed on these terms in order for the flatness of the
inflationary potential is maintained. On the other hand, some of
these terms assist us to precisely reconcile the resulting scalar
spectral index with the current PLANCK measurements while the
other inflationary observables are in agreement with data.
\\ \\ {\sl\bfseries
Published in}~~{PoS  Corfu {\bf 2012}, 061 (2013)}. }

\FullConference{Proceedings of the Corfu Summer Institute 2012 \\
                 September 8-27, 2012\\
                 Corfu, Greece}

\begin{document}

\section{Introduction}

\emph{Non-minimal (chaotic) inflation} ({\sf\small nMI}) is a
class of inflationary models which arises in the presence of a
strong enough non-minimal coupling function between the inflaton
field and the Ricci scalar curvature. In this talk, which is based
on Refs.~\cite{nmi,nmN,nmH, nMCI}, we first briefly review the
basic ingredients of nMI in a non-\emph{Supersymmetric}
({\sf\small SUSY}) framework (Sec.~\ref{Fhi}) and constrain the
parameters of two typical models in Sec.~\ref{nmi} taking into
account the observational requirements described in \Sref{obs}.
Throughout the text, the subscript $,\chi$ denotes derivation
\emph{with respect to} (w.r.t) the field $\chi$ and we follow the
conventions of \cref{nmi}.

\subsection{Coupling non-Minimally the Inflaton to Gravity}\label{Fhi}

The action of an inflaton $\sigma$ with potential $V_{\rm I}
\left(\phi\right)$  non-minimally coupled to Ricci scalar $\rcc$
through a coupling function $\fr(\sigma)$, in the \emph{Jordan
frame} (JF), takes the form:
\beq \label{action1} {\sf  S} = \int d^4 x \sqrt{-\mathfrak{g}}
\left(-\frac{1}{2} \mP^2\fr(\sigma)\rcc
+\frac{\fkk(\phi)}{2}g^{\mu\nu}
\partial_\mu \sg\partial_\nu \sg- V_{\rm I} \left(\sigma\right)\right),
\eeq
where $\mP = 2.44\cdot 10^{18}~\GeV$  is the reduced Planck mass,
$\mathfrak{g}$ is the determinant of the background
Friedmann-Robertson-Walker metric, $g^{\mu\nu}$. We allow also for
a kinetic mixing through the function $\fkk(\phi)$. We can write
${\sf S}$ in the \emph{Einstein frame} ({\sf\small EF}) as follows
\beq {\sf  S}= \!\int d^4 x
\sqrt{-\what{\mathfrak{g}}}\left(-\frac12 \mP^2
\rce+\frac12\geu^{\mu\nu} \partial_\mu \se\partial_\nu
\se-\Ve_{\rm I}(\se)\right)\,, \label{action} \eeq
by performing a conformal transformation \cite{nmi} according to
which we define the EF metric
\beq\label{contr}
\geu_{\mu\nu}=\fr\,g_{\mu\nu}~~\Rightarrow~~\left\{\bem
\sqrt{-\what{\mathfrak{g}}}=\fr^2\sqrt{-\mathfrak{g}}\hspace*{0.2cm}\mbox{and}\hspace*{0.2cm}
\geu^{\mu\nu}=g^{\mu\nu}/\fr\,, \hfill \cr
\widehat\rcc=\left(\rcc+3\Box\ln \fr+3g^{\mu\nu} \partial_\mu
\fr\partial_\nu \fr/2\fr^2\right)/\fr\,, \hfill \cr\eem
\right.\eeq
where $\Box=\lf
-\mathfrak{g}\rg^{-1/2}\partial_\mu\lf\sqrt{-\mathfrak{g}}\partial^\mu\rg$
and hat is used to denote quantities defined in the EF. We also
introduce the EF canonically normalized field, $\se$, and
potential, $\Ve_{\rm I}$, defined as follows:
\beq \label{VJe}
\left(\frac{d\se}{d\sigma}\right)^2=J^2=\frac{\fk}{\fr}+{3\over2}\mP^2\left({f_{{R},\sg}\over
\fr}\right)^2~~\mbox{and}~~\Ve(\se) = \frac{V_{\rm
I}\lf\se(\sigma)\rg}{\fr\lf\se(\sigma)\rg^2}\,\cdot\eeq
For convenient choices of $V_{\rm I}(\phi)$ and $\fr(\phi)$ we can
obtain a sufficiently flat $\Ve_{\rm I}(\se)$ which can support
nMI. The analysis of nMI in the EF using the standard slow-roll
approximation is equivalent with the analysis in JF. We have just
to keep in mind the dependence of $\what\phi$ on $\phi$.

\subsection{Inflationary Observables -- Constraints} \label{obs}

Under the assumptions that {\sf (i)} the curvature perturbation
generated by $\sigma$ is solely responsible for the observed one
and {\sf (ii)} nMI is followed in turn by a decaying-inflaton,
radiation and matter domination, the parameters of nMI can be
restricted by requiring that:

\begin{figure}[t]\vspace*{-0.45cm}
\begin{minipage}{70mm}
\includegraphics[height=3in,angle=-90]{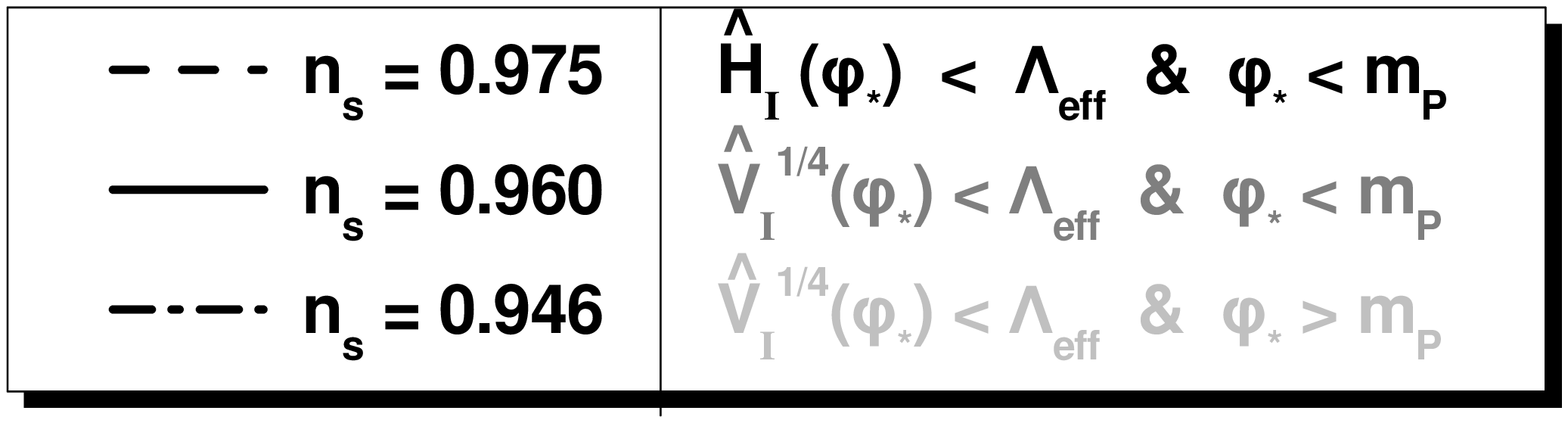}
\end{minipage}
\hfil \hspace*{1.5cm}\begin{minipage}{65mm} \caption{\sl
Conventions for the type and the color of the lines used for the
various restrictions on the parameters of our models.}\label{fig0}
\end{minipage}

\end{figure}

\paragraph{\bf (i)} The number of e-foldings, $\widehat N_{*}$, that
the scale $k_*=0.05/{\rm Mpc}$ suffers during nMI, leads to a
solution of the horizon and flatness problems of standard big
bang, i.e., \cite{plin}
\begin{equation}
\label{Nhi}  \int_{\se_{\rm f}}^{\se_{*}}\, \frac{d\se}{m^2_{\rm
P}}\: \frac{\Ve_{\rm I}}{\Ve_{\rm I,\se}}= \int_{\sigma_{\rm
f}}^{\sigma_{*}}\, J^2\frac{\Ve_{\rm I}}{\Ve_{\rm
I,\sigma}}{d\sigma\over\mP^2}\: \simeq19.4+2\ln{V_{\rm
I}(\sg_{*})^{1/4}\over{1~{\rm GeV}}}-{4\over 3}\ln{V_{\rm
I}(\sg_{\rm f})^{1/4}\over{1~{\rm GeV}}}+ {1\over3}\ln {T_{\rm
rh}\over{1~{\rm GeV}}}+{1\over2}\ln{\fr(\sg_{\rm f})\over
\fr(\sg_*)},
\end{equation}
where $\sg_{\rm f}~[\se_{\rm f}]$ is the value of $\sg~[\se]$ at
the end of nMI, which can be found, in the slow-roll approximation
and for the considered in this paper models, from the condition
$$ {\sf max}\{\widehat\epsilon(\sigma_{\rm
f}),|\widehat\eta(\sigma_{\rm f})|\}=1,~~~\mbox{where}$$
\beq \label{sr}\widehat\epsilon= {\mP^2\over2}\left(\frac{\Ve_{\rm
I,\se}}{\Ve_{\rm I}}\right)^2={\mP^2\over2J^2}\left(\frac{\Ve_{\rm
I,\sigma}}{\Ve_{\rm I}}\right)^2 ~~\mbox{and}~~\widehat\eta=
m^2_{\rm P}~\frac{\Ve_{\rm I,\se\se}}{\Ve_{\rm I}}={\mP^2\over
J^2}\left(\frac{\Ve_{\rm I,\sigma\sigma}}{\Ve_{\rm
I}}-\frac{\Ve_{\rm I,\sigma}}{\Ve_{\rm I}}{J_{,\sg}\over
J}\right)\cdot \eeq
Also $\Trh$ is the reheat temperature after nMI, which is taken
$\Trh=10^8~\GeV$ throughout, since its variation over some orders
of magnitude influences only weakly $\Ne_*$, which remains close
to $50$.

\paragraph{\bf (ii)} The amplitude $A_{\rm s}$ of the power spectrum of the curvature perturbation
generated by $\sigma$ at the pivot scale $k_{*}$ is consistent
with data~\cite{plin}
\begin{equation}  \label{Prob}
\sqrt{A_{\rm s}}=\: \frac{1}{2\sqrt{3}\, \pi\mP^3} \;
\frac{\Ve_{\rm I}(\sex)^{3/2}}{|\Ve_{\rm
I,\se}(\sex)|}=\frac{|J(\sigma_*)|}{2\sqrt{3}\, \pi\mP^3} \;
\frac{\Ve_{\rm I}(\sigma_*)^{3/2}}{|\Ve_{\rm
I,\sigma}(\sigma_*)|}\simeq4.685\cdot 10^{-5},
\end{equation}
where $\sigma_*~[\se_*]$ is the value of $\sg~[\se]$ when $k_*$
crosses outside the inflationary horizon.

\paragraph{\bf (iii)}  The (scalar) spectral index, $n_{\rm s}$, its
running, $a_{\rm s}$, and the scalar-to-tensor ratio $r$ --
estimated through the relations:
\beq\label{ns} n_{\rm s}=\: 1-6\widehat\epsilon_*\ +\
2\widehat\eta_*,\hspace*{0.2cm} \alpha_{\rm s}
=\:2\left(4\widehat\eta_*^2-(n_{\rm
s}-1)^2\right)/3-2\widehat\xi_*\hspace*{0.2cm}
\mbox{and}\hspace*{0.2cm} r=16\widehat\epsilon, \eeq
where $\widehat\xi=\mP^4 {\Ve_{\rm I,\widehat\sigma} \Ve_{\rm
I,\widehat\sigma\widehat\sigma\widehat\sigma}/\Ve_{}^2}=
\mP^2\,\Ve_{\rm I,\sigma}\,\widehat\eta_{,\sigma}/\Ve_{\rm
I}\,J^2+2\widehat\eta\widehat\epsilon$ and the variables with
subscript $*$ are evaluated at $\sigma=\sigma_{*}$ -- are
consistent  with the fitting of the data \cite{plin} with
$\Lambda$CDM model, i.e.,
\begin{equation}\label{nswmap}
\mbox{\ftn\sf (a)}~~ n_{\rm s}=0.9603\pm0.0146,~~\mbox{\ftn\sf
(b)}~~ -0.0314\leq a_{\rm s}\leq0.0046 ~~\mbox{and}~~\mbox{\ftn\sf
(c)}~~r<0.135~~~\mbox{at 95$\%$ c.l.}\end{equation}

\paragraph{\bf (iv)} The effective theory describing nMI remains valid.
This aim can be achieved if the ultraviolet cut-off scale
\cite{cutoff}, $\Ldf$, remains larger than the inflationary scale
which is represented by $\Ve_{\rm I}(\sg_*)^{1/4}$ or, less
restrictively, by the corresponding Hubble parameter, $\widehat
H_{\rm I*}=\Ve_{\rm I}(\sg_*)^{1/2}/\sqrt{3}\mP$, i.e.
\beq \label{Vl}\mbox{\ftn\sf (a)}~~ \Ve_{\rm
I}(\sg_*)^{1/4}\leq\Ldf~~\mbox{or}~~\mbox{\ftn\sf (b)}~~\widehat
H_{\rm I*}\leq\Ldf,\eeq
where assuming $\fr=1+\ck\lf\sg/\mP\rg^n$ we obtain \cite{riotto}
$\Ldf\simeq\mP$ for $n=1$ and $n>7/3$ and $\Ldf\simeq\mP/\ck$ for
for $1<n\leq7/3$. Note that the imposition of \Eref{Vl} implies
automatically the validity of $\Ve_{\rm I}(\sg_*)^{1/4}\leq\mP$
which is necessary in order to avoid possible corrections from
quantum gravity. Less favored, from theoretical point of view, is
the domain of parameters where $\sg_*\geq\mP$. The conventions
adopted for the description of the various restrictions on the
parameters of our models are shown in \Fref{fig0}.

\subsection{Non-SUSY Models of nMI}\label{nmi}

Using the criteria of \Sref{obs}, it would be instructive to test
two simple models of inflation based on quartic (\Sref{nmi1}) and
quadratic (\Sref{nmi2}) potential using minimal and, conveniently
chosen, non-minimal coupling functions.

\subsubsection{The Quartic Potential}\label{nmi1}

Adopting the potential $V_{\rm I}=\ld\sig^4/4$ for the inflaton
$\phi$ we can analyze the cases:

\paragraph{\bf (i)} If we use $\fr=1$ and $\fkk=1$, i.e. the minimal coupling to
gravity, the slow-roll parameters and the number of $e$-foldings,
suffered from $k_*$ during \emph{minimal inflation} {\ftn\sf (MI)}
can be calculated applying \Eref{sr} and \Eref{Nhi} -- after
removing hats and setting $J=1$ -- with results
\beqs\beq\epsilon\simeq {8\mP^2/\sig^2},~~
\eta\simeq{12\mP^2/\sig^2}\>\>\>\mbox{and}\>\>\>N_{*}\simeq{\sig_*^2/
8\mP^2}\cdot\eeq
Therefore, the values of $\sig$ at the end of MI and the horizon
crossing of $k_*$ are transplanckian, since
\beq
\sigf={2\sqrt{3}\mP}\>\>\>\mbox{and}\>\>\>\>\sig_*=2\sqrt{2N_{*}}\mP.\eeq
Moreover, the normalization of \Eref{Prob} imposes the condition
\beq  \sqrt{A_{\rm s}}
\simeq{\sqrt{\ld}\sig_*^3\over16\sqrt{3}\pi\mP^3}=4.685\cdot10^{-5}\>
\Rightarrow\> \ld^{1/2}\simeq{\sqrt{3\over2}}4.685\cdot10^{-5}\pi
N_*^{-3/2}\> \Rightarrow\> \ld\simeq2\cdot10^{-13}\label{lAs}\eeq
for $N_*\simeq53$. This value of $\ld$ signals an ugly tuning.
From \Eref{ns} we get
\beq n_{\rm s}\simeq1-{3/N_*}\simeq0.947, \>\>\> \alpha_{\rm s}
\simeq-{3/N_*^2}\simeq9.5\cdot10^{-4} \>\>\> \mbox{and}\>\>\>
r\simeq{16/N_*}\simeq0.28,\eeq\eeqs
with the last value being in clear contradiction with
\sEref{nswmap}{c}.

\paragraph{\bf (ii)} If we employ $\fr(\sig)=1+\ck(\sig/\mP)^2$ and $\fkk=1$,
i.e., the standard non-minimal coupling to gravity, we find from
\Eref{VJe}, for $\ck\gg1$
\beqs\beq
J\simeq{\sqrt{6}\mP/\sg}\>\>\>\mbox{and}\>\>\>\Vhi={\ld\phi^4/4\fr^2}\simeq
\ld\mP^4/4\ck^2.\>\>\eeq
We observe that $\Vhi$ exhibits an almost flat plateau. From
\eqs{Nhi}{sr} we find
\beq\widehat\epsilon\simeq {4\mP^4/3\ck^2\sig^4},\>\>\>
\widehat\eta\simeq-{4\mP^2/3\ck\sig^2}\>\>\>\mbox{and}\>\>\>\widehat{N}_{*}\simeq{3\ck\sig_*^2/4\mP^2}\,.\eeq
Therefore, $\sigf$ and $\sig_*$ are found from the condition of
\Eref{sr} and the last equality above, as follows
\beq
\label{sigs1}\sigf=\sqrt[4]{4/3}{\mP/\sqrt{\ck}}~~~\mbox{and}~~~
\sig_*=2\mP\sqrt{\widehat N_*/3\ck}.\eeq
Also the normalization of \Eref{Prob} implies the following
relation between $\ck$ and $\sqrt{\ld}$
\beq \label{Prob1}\sqrt{A_{\rm s}}\simeq
{\sqrt{\ld}\widehat{N}_{*}\over6\sqrt{2}\pi\ck}=4.685\cdot10^{-5}
\>\Rightarrow\> \ck\simeq4.2\cdot10^4\, \sqrt{\ld}\eeq
for $\Ne_*\simeq52$. Obviously $\ld$ can be larger than that in
\Eref{lAs}. From \Eref{ns} we get
\beq \label{nswmap1} n_{\rm s}\simeq1-{2/\widehat N_*}\simeq0.965,
\>\>\> \alpha_{\rm s} \simeq{-2/\widehat
N^2_*}\simeq-6.4\cdot10^{-4}\>\>\> \mbox{and}\>\>\>
r\simeq{12/\widehat N^2_*}\simeq4\cdot 10^{-3},\eeq
which are in agreement with \Eref{nswmap}. If, in addition, we
impose the requirement $\phi\leq\mP$ and that of \sEref{Vl}{a} we
end-up with following ranges -- c.f. \cref{nmi}:
\beq
\label{res4}74\simeq4\Ne/3\lesssim\ck\lesssim300~~\mbox{and}~~
0.3\lesssim\lambda/10^{-5}\lesssim4.6\,.\eeq\eeqs
Therefore, the presence of $\fr$ can rescue the model based on the
simplest quartic potential.

\subsubsection{The Quadratic Potential}\label{nmi2}

Focusing on the potential $V_{\rm I}=m^2\sig^2/2$ for the inflaton
$\phi$ we concentrate on the cases:

\paragraph{\bf (i)} If we use $\fr=1$ and $\fkk=1$, i.e. the minimal coupling to
gravity, and work along the lines of \Sref{nmi1}, from
\eqs{sr}{Nhi} we find
\beqs\beq \label{mci1}\epsilon=\eta=2\mP^2/\sg^2\hspace*{0.2cm}
\mbox{and}\hspace*{0.2cm}N_*=\lf\sg_*^2-\sg_{\rm
f}^2\rg/4\mP^2.\eeq
Imposing the condition of \Eref{sr} and solving the second
expression above w.r.t $\sg_*$ we find
\beq \sg_{\rm
f}=\sqrt{2}\mP~~\mbox{and}~~\sg_*\simeq2\mP\sqrt{N_*}. \eeq
As in \Sref{nmi1} transplanckian $\sg$'s are required. Enforcing
\Eref{Prob}, we extract
\beq \sqrt{A_{\rm s}}
\simeq{m\sig_*^2\over4\sqrt{6}\pi\mP^3}=4.686\cdot10^{-5}\>
\Rightarrow\> m\simeq4.686\cdot10^{-5}\sqrt{6}\pi\mP N_*^{-1}\>
\Rightarrow\> m\simeq10^{13}~\GeV,\eeq
which can be related to the masses of the right-handed neutrinos
\cite{nmN}. Applying \Eref{ns} we get
\beq \ns\simeq1-{2/ N_*}\simeq0.965,~~\as\simeq-{2/
N^2_*}\simeq-6.5\cdot10^{-4}~~\mbox{and}~~
r\simeq8/N_*\simeq0.13.\label{mci2}\eeq\eeqs
Although $\ns$ is close to the observationally favored one, $r$ is
in tension with \sEref{nswmap}{c}.

\paragraph{\bf (ii)} If we employ $\fr(\sig)=1+\ck\sig/\mP$ and $\fkk=1$,
i.e. a linear non-minimal coupling to gravity, we find from
\Eref{VJe}, for $\ck\gg1$
\beqs\beq\label{nmci1b} J\simeq\sqrt{3/2}\mP
\sg^{-1}~~\mbox{and}~~\Vhi\simeq {m^2\mP^2/2\ck^2}\,,\eeq
where a plateau is again generated as in \Sref{nmi1}. Employing
\eqss{Nhi}{sr}{nmci1b}, the slow roll parameters and
$\widehat{N}_*$ read
\beq\label{nmci2b} \widehat\epsilon
\simeq{4\mP^2/3\ck^2\sg^2},~~\widehat\eta \simeq-{4\mP/3\ck\sg}
~~\mbox{and}~~\widehat{N}_{*}\simeq {3\ck}{\sg_*}/4\mP.\eeq
Imposing the condition of \Eref{sr} and solving then the latter
equation w.r.t $\sg_*$ we arrive at
\beq\label{nmci4b}{\sg_{\rm
f}}\simeq2\mP/\sqrt{3}\ck\hspace*{0.2cm}\mbox{and}\hspace*{0.2cm}
{\sg_*}\simeq4\mP\widehat{N}_*/3\ck\cdot \eeq
On the other hand, \Eref{Prob} implies a relation between $m$ and
$\ck$ -- cf. \Eref{Prob1}
\beq \label{Prob2}\sqrt{A_{\rm s}}\simeq
{m\widehat{N}_{*}\over6\pi\mP\ck}=4.686\cdot10^{-5}
\>\Rightarrow\>m=4.13\cdot10^{13}\ck~\GeV.\eeq
Applying \Eref{ns} we find that the predictions of the model w.r.t
$\ns, \as$ and $r$ are identical to those in \Eref{nswmap1}. If,
in addition, we impose the requirement $\phi\leq\mP$ and that of
\sEref{Vl}{a} we end-up with following ranges -- c.f. \cref{nmi}:
\beq \label{res2}\ck\gtrsim77\simeq
4\Ne_*/3~~\mbox{and}~~m\gtrsim2.9\cdot10^{15}~\GeV.\eeq\eeqs
Note that \Eref{Vl} does not constrain the parameters in sharp
contrast to \Eref{res4}. Therefore, the presence of a linear $\fr$
renders observationally more interesting the model based on the
simplest quadratic potential without to cause problems with the
perturbative unitarity \cite{cutoff,riotto}.

\subsection{Outline}\label{plan}

It would be certainly interesting to enquire if it is possible to
realize  similar models of nMI in a SUSY framework where the
hierarchy problem of \emph{Grand Unified Theories} ({\sf\small
GUTs}) is elegantly resolved. We below describe the formulation of
nMI in the context of \emph{Supergravity} ({\sf\small SUGRA}) and
we specify three models of nMI: two with a gauge singlet inflaton
(coupled to another gauge singlet in a linear-quadratic or a
bilinear superpotential term) and one with a gauge non-singlet
inflaton.

\section{Realization of nMI Within SUGRA}\label{fhim}

In \Sref{sugra1} we present the basic formulation of a theory
which exhibits non-minimal coupling of scalar fields to $R$ within
SUGRA and in \Sref{sugra2} we outline our strategy in constructing
viable models of nMI.

\subsection{The General Set-up} \label{sugra1}

Our starting point is the EF action for the scalar fields
$\Phi^\al$ within SUGRA \cite{linde1,nmN} which can be written as
\beqs \beq\label{Saction1} {\sf S}=\int d^4x \sqrt{-\what{
\mathfrak{g}}}\lf-\frac{1}{2}\mP^2 \rce +K_{\al\bbet}\geu^{\mu\nu}
D_\mu\Phi^\al D_\nu\Phi^{*\bbet}-\Ve\rg, \eeq
where the following notation is adopted
\beq \label{Kab}
K_{\al\bbet}={\Khi_{,\Phi^\al\Phi^{*\bbet}}}>0\>\>\>\mbox{and}\>\>\>D_\mu\Phi^\al=\partial_\mu\Phi^\al-A^A_\mu
k^\al_A\eeq
are the covariant derivatives for scalar fields $\Phi^\al$. Here
and henceforth the scalar components of the various superfields
are denoted by the same superfield symbol. Also $A^A_\mu$ stand
for the vector gauge fields and $k^\al_A$ is the Killing vector,
defining the gauge transformations of the scalars \cite{linde1}.
The EF potential, $\Ve$, is given in terms of the K\"ahler
potential, $K$, and the superpotential, $W$, by
\beq \Ve=\Ve_{\rm F}+ \Ve_{\rm D}\>\>\>\mbox{with}\>\>\> \Ve_{\rm
F}=e^{\Khi/\mP^2}\left(K^{\al\bbet}{\rm F}_\al {\rm
F}^*_\bbet-3\frac{\vert W\vert^2}{\mP^2}\right)
\>\>\>\mbox{and}\>\>\>\Ve_{\rm D}= {1\over2}g^2 \sum_a {\rm D}_a
{\rm D}_a. \label{Vsugra} \eeq
Here, $g$ is the unified gauge coupling constant and the summation
is applied over the generators $T_a$ of a considered gauge group
-- a trivial gauge kinetic function is adopted. Also we use the
shorthand
\beq \label{Kinv} K^{\bbet\al}K_{\al\bar
\gamma}=\delta^\bbet_{\bar \gamma},\>\>{\rm F}_\al=W_{,\Phi^\al}
+K_{,\Phi^\al}W/\mP^2~~\mbox{and}~~{\rm D}_a=\Phi_\al\lf
T_a\rg^\al_\bt K^\bt~~\mbox{with}~~
K^{\al}={\Khi_{,\Phi^\al}}.\eeq\eeqs

By performing a conformal transformation and adopting a frame
function $\Omega$ which is related to $K$ as follows
\beq-\Omega/3
=e^{-K/3\mP^2}\>\Rightarrow\>K=-3\mP^2\ln\lf-\Omega/3\rg,\label{Omg1}\eeq
we arrive at the following action
\beq {\sf S}=\int d^4x
\sqrt{-\mathfrak{g}}\lf-\frac{\mP^2}{2}\lf-{\Omega\over3}\rg
\rcc+\mP^2\Omega_{\al{\bbet}}D_\mu\Phi^\al
D^\mu\Phi^{*\bbet}-\Omega {\cal A}_\mu{\cal A}^\mu/\mP^2-V
 \rg, \label{Sfinal}\eeq
where $g_{\mu\nu}=-\lf 3/\Omega\rg\geu_{\mu\nu}$ is the JF metric,
we use the shorthand notation $\Omega_\al=\Omega_{,\Phi^\al}$ and
$\Omega_\aal=\Omega_{,\Phi^{*\aal}}$ and ${\cal A}_\mu$ is the
purely bosonic part of the on-shell value of the auxiliary field
$A_\mu$ given by
\beq {\cal A}_\mu =-i\mP^2\lf
D_\mu\Phi^\al\Omega_\al-D_\mu\Phi^{*\aal}\Omega_\aal\rg/2\Omega\,.
\label{Acal}\eeq
It is clear from \Eref{Sfinal} that ${\sf S}$ exhibits non-minimal
couplings of the $\Phi^\al$'s to $\rcc$. However, $\Omega$ enters
the kinetic terms of the $\Phi^\al$'s too. In general, $\Omega$
can be written as \cite{linde1}
\beq -\Omega/3=1 - F_K\lf\Phi^\al {\Phi}^{*\aal}\rg/3 +\big(
\Fcr(\Phi^\al)+{\Fcr}^*(\Phi^{*\aal})\big), \label{Omg}\eeq
where $F_K$ is a dimensionless real function while $\Fcr$ is a
dimensionless, holomorphic function. For $\Fcr>F_K$, $F_K$
expresses mainly the kinetic terms of the $\Phi^\al$'s whereas
$\Fcr$ represents the non-minimal coupling to gravity -- note that
$\Omega_{\al{\bbet}}$ is independent of $\Fcr$ since
$F_{R,\Phi^\al\Phi^{*\bbet}}=0$. In order to get canonical kinetic
terms, we need \cite{linde2} ${\cal A}_\mu=0$ and
$F_{K\al{\bbet}}\simeq\delta_{\al{\bbet}}$. The first condition is
attained when the dynamics of the $\Phi^\al$'s is dominated only
by the real moduli $|\Phi^\al|$. The second condition is satisfied
by the choice
\beq \label{Fkdef} F_K\lf|\Phi^\al|^2 \rg=
{|\Phi^\al|^2/\mP^2}\,+\, k_{\al\bt}\,
{|\Phi^\al|^2|\Phi^\bt|^2/\mP^4}\eeq
with sufficiently small coefficients $k_{\al\bt}$. Here we assume
that the $\Phi$'s are charged under a global -- see
Secs.~\ref{fhi3} and \ref{fhi2} -- or gauge -- see \Sref{fhi1} --
$U(1)$ symmetry, so as mixed terms of the form
$\Phi^\al{\Phi}^*_{\bt}$ are disallowed. The inclusion of the
fourth order term for the accompanying non-inflaton field,
$\Phi^1:=S$ is obligatory in order to evade \cite{linde1} a
tachyonic instability occurring along this direction. As a
consequence, all the allowed terms are to be considered in the
analysis for consistency.

\subsection{Modeling nMI in SUGRA}\label{sugra2}

The realization of nMI in SUGRA requires $\what V_{\rm D}=0$. This
condition may be attained, when  the inflaton is (the radial part
of) a gauge:
\begin{itemize}
\item Singlet, by introducing an extra field $\Phi^2:=\Phi$ which
obviously has zero contribution to $\what V_{\rm D}$.

\item Non-singlet by introducing a conjugate pair of Higgs
superfields, $\Phi^2:=\Phi$ and $\Phi^3:=\bar\Phi$, which are
parameterized so as $\what V_{\rm D}=0$.
\end{itemize}

The presence of $S$ is crucial for the realization of our
scenaria, since it assists us to isolate via its derivative the
contribution of the inflaton(s) in $\Ve$, \Eref{Vsugra}. Indeed,
placing $S$ at the origin the resulting $\Ve=\Vhio$ -- in both
cases above -- is equal to
\beq \Vhio= e^{K/\mP^2}K^{SS^*}\, W_{,S}\, W^*_{,S^*}={V_{\rm
F}/\fsp\fr^2}~~\mbox{since}~~e^{K/\mP^2}={1/\fr^3}
~~\mbox{and}~~K^{SS^*}={\fr/\fsp},\label{Vhig}\eeq
where $\fr=-\Ohi/3$ and $\fsp=\mP^2\Omega_{,SS^*}$. Also $V_{\rm
F}=\left|W_{,S}\right|^2$ is the F-term SUSY potential, obtained
from \Eref{Vsugra} with D$^\al=0$ in the limit $\mP\to\infty$.
Given that $\fsp\ll\fr$, the construction of an inflationary
plateau is reduced in the selection of the appropriate $W$ and
$\Fcr$ so that $\Vhio\simeq V_{\rm F}/\fr^2$ is almost constant.
In the case of a gauge-singlet inflaton, where D$^\al=0$, this
objective can be achieved with one of the two choices:

\begin{itemize}

\item If we set \cite{nmN,linde2} $W=\ld S\Phi^2$ and
$\Fcr=1+\ck\Phi^2/\mP^2$ -- cf. \Sref{nmi1} --, we obtain $V_{\rm
F}=\ld^2|\Phi|^4$ and $\fr^2\simeq\ck^2|\Phi|^4$ for $\ck\gg1$.
Therefore, \Eref{Vhig} implies that $\Vhio$ turns out to be almost
constant.

\item If we choose \cite{nMCI} $W=mS\Phi$ and $\Fcr=1+\ck\Phi/\mP$
-- cf. \Sref{nmi2} --, we find $V_{\rm F}=m^2|\Phi|^2$ and
$\fr^2\simeq\ck^2|\Phi|^2$ for $\ck\gg1$. \Eref{Vhig} gives again
an almost constant $\Vhio$.

\end{itemize}
In the case of a gauge non-singlet inflaton, we can take
\cite{nmH} $W=\ld S\lf\Phi\bar\Phi-M^2\rg$ which can be combined
with the gauge invariant $\Fcr$, $\Fcr=1+\ck\bar\Phi\Phi/\mP^2$ --
cf. \cref{jones2}. For $|\Phi|=|\bar\Phi|$, we expect $V_{\rm
D}=0$, $V_{\rm F}\simeq\ld^2|\Phi|^4$ (for $|\Phi|\gg M$) and
$\fr^2\simeq\ck^2|\Phi|^4$ for $\ck\gg1$. Therefore, \Eref{Vhig}
again implies that $\Vhio$ is almost constant. The GUT gauge group
is spontaneously broken during and at the end of nMI, where $\Phi$
and $\bar\Phi$ acquire their \emph{vacuum expectation values}
({\sf\ftn VEVs}) $\vev{\Phi}=\vev{\bar\Phi}=M$ with SUSY unbroken
-- up to tiny corrections from SUSY breaking effects.

In the following we show details on the realization of these three
scenaria.  Our analysis is carried out exclusively in the EF
substituting into \Eref{Vsugra}, \eqs{Omg1}{Omg} and the chosen
$\Fcr$. Then, \Eref{VJe} can be employed in order to find the
canonically normalized inflaton field $\se$ in terms of $\sg$
defining
\beq \label{fkdef} {\fk}=\mP^2\Omega_{\al{\bbet}}\dot\Phi^\al
\dot\Phi^{*\bbet}/\lf\dot\phi^2/2\rg~~\mbox{and}~~\fr=-\Ohi/3.\eeq
We also check the stability of the inflationary trajectory w.r.t
the fluctuations of the non-inflaton fields and find the whole
spectrum during nMI. Employing the derived masses we find the
one-loop radiative correction $\Delta V$ and the corresponding EF
potential using the Coleman-Weinberg formula
\beq \Vhi=\Vhio+\Delta V\>\>\>\mbox{with}\>\>\>\Delta
V=\frac{1}{64\pi^2}\sum_i(-)^{{\rm F}_i}M_i^4\ln
\frac{M_i^2}{\Lambda^2}\,,\label{Vhic} \eeq
where the sum extends over all helicity states $i$, ${\rm F}_i$
and $M_i^2$ is the fermion number and mass squared of the $i$th
state and $\Lambda$ is a renormalization mass scale.

\section{nMI with Quartic Potential for a Gauge Singlet Inflaton}\label{fhi3}

This setting is realized in the presence of two superfields $S$
and $\Phi$ charged under a global $U(1)$ symmetry with charges
$2$ and $-1$ respectively. In particular we take
\beq \label{mfhi3} W=\ld S\Phi^2,~\Fcr={c_\Phi\over
4\mP^2}\Phi^2,~
\Fk={|S|^2\over\mP^2}+{|\Phi|^2\over\mP^2}-2\ks{|S|^4\over\mP^4}-
2\kpp{|\Phi|^4\over\mP^4}-2\ksp {|S|^2|\Phi|^2\over\mP^4}\cdot\eeq
In \Sref{fhi31} we describe the salient features of this model and
in \Sref{num3} we expose our results.

\subsection{Structure of the Inflationary
Potential}\label{fhi31}

The EF F--term (tree level) SUGRA scalar potential, $\Ve$, of this
model is obtained from \Eref{Vsugra} upon substitution of
\Eref{mfhi3} into \eqs{Omg1}{Omg} with $\Phi^1:=S$ and
$\Phi^2:=\Phi$ -- note that $D^\al$ vanish by construction. We can
verify that along the trajectory
\beq S=0\>\>\>\mbox{and}\>\>\>\th:={\sf arg}\Phi=0
\label{inftr3}\eeq
and for $c_\Phi\gg1$, $\Ve$ develops a plateau with almost
constant potential energy density, $\Vhio$ and corresponding
Hubble parameter $\He_{\rm I}$ found by \Eref{Vhig}:
\beqs\beq\label{3Vhio} \Vhio=
{\ld^2\sg^4\over4\fsp\fr^2}\simeq{\ld^2\mP^4\over4\fsp\ck^2}\>\>\>\mbox{and}\>\>\>
\He_{\rm
I}={\Vhio^{1/2}\over\sqrt{3}\mP}\simeq{\ld\mP\over2\sqrt{3}\fsp\ck}\,,\eeq
where $\fr$ and $\fsp$ are calculated by employing their
definitions below \Eref{Vhig} \beq
\fr=1+\ck\xsg^2+{\kpp\xsg^4/6}\>\>\>\mbox{and}\>\>\>\fsp=1-\ksp\xsg^2~~
\mbox{with}~~\xsg=\sg/\mP~~\mbox{and}~~\ck=c_{\Phi}/4-1/6.\label{fs3}
\eeq\eeqs Expanding $\Phi$ and $S$ in real and imaginary parts
according to the prescription
\beq \Phi={\sg e^{i\th}/\sqrt{2}}\>\>\>\mbox{and}\>\>\>S=
{(s_1+is_2)/\sqrt{2}},\label{cannor3a} \eeq
we find that along the trajectory of \Eref{inftr3}, $K_{\al\bbet}$
defined in \Eref{Kab} takes the form
\bea \lf K_{\al\bbet}\rg=\diag\lf
J^2,\fsp/\fr\rg\>\>\>\mbox{where}\>\>\> J\simeq
{\sqrt{3}\ck^2\sg^2/\sqrt{2} \fr^2}\simeq
{\sqrt{6}/\xsg},\>\>\label{VJe3}\eea
as can be inferred from \Eref{VJe} for $\fk=1-4\kpp\xsg^2$ -- see
\Eref{fkdef}. Consequently, we can introduce the EF canonically
normalized fields, $\se, \widehat \th$ and $\widehat s_i$, with
$i=1,2$ as follows -- cf. \cref{linde1,linde2,nmN,nmH}:
\beq \label{K3}K_{\al\bbet}\dot\Phi^\al
\dot\Phi^{*\bbet}=\frac12\lf\dot{\what\phi}^{2}+\dot{\what
\th}^{2}+\dot{\what s}_1^2+\dot{\what s}_2^2\rg,\eeq
where the dot denotes derivation w.r.t the JF cosmic time and the
hatted fields are defined as follows
\beq  \label{cannor3b} {d\widehat \sg/d\sg}=J,\>\>\> \widehat
\th\simeq J\sg \th \>\>\>\mbox{and}\>\>\>\what s_i=
\sqrt{{\fsp/\fr}}\,s_i \>\>\>\mbox{with}\>\>\>i=1,2.\eeq

\renewcommand{\arraystretch}{1.5}

\begin{table}[!t]
\bec\begin{tabular}{|c|c|l|}\hline
{\sc Fields} &{\sc Eingestates} & \hspace*{3.cm}{\sc Mass Squared}\\
\hline \hline
$1$ real scalar &$\what \th$ & $m^2_{\what \th}=\ld^2\lf1+6\ck\rg\mP^2\xsg^4/3\fsp\fr^3J^2\simeq4\He_{\rm I}^2$\\
$2$ real scalars &$\what s_1,~\what s_2$ & $m^2_{\what
s}\simeq\ld^2\mP^2\lf2+\ck^2
(12\ks(1+\ck\xsg^2)-1)\xsg^4\rg/6\ck^2\fsp^3\fr^2$\\\hline
$2$ Weyl spinors & $\what \psi_\pm={\what{\psi}_{\Phi}\pm
\what{\psi}_{S}\over\sqrt{2}}$& $m^2_{\what
\psi\pm}\simeq\ld^2\mP^2(2-\ksp\xsg^2+\ksp\ck\xsg^4)^2/18\fsp^3\fr^2$\\
\hline
\end{tabular}\eec
\caption{\sl The mass spectrum of the model along the inflationary
trajectory of Eq.~(3.2).}\label{tab3}
\end{table}

To check the stability of $\Ve$ in \Eref{Vsugra} along the
trajectory in \Eref{inftr3} w.r.t the fluctuations of $\what\th$
and $\what s_i$ we construct the mass spectrum of the theory. Our
results are summarized in \Tref{tab3}. From there it is evident
that $\ks\gtrsim0.5$ assists us to achieve $m^2_{\what{s}}>0$. We
have also numerically verified that the various masses remain
greater than $\Hhi$ during the last $50$ e-foldings of nMI, and so
any inflationary perturbations of the fields other than the
inflaton are safely eliminated. In \Tref{tab3} we also present the
masses squared of chiral fermions along the direction of
\Eref{inftr}, which can be served for the calculation of $\Delta
V$ in \Eref{Vhic}. We observe that the fermionic (4) and bosonic
(4) \emph{degrees of freedom} (d.o.f) are equal -- here we take
into account the 1 d.o.f of $\what\phi$ which is not perturbed.
$\Delta V$ has no impact on our results, since the slope of the
inflationary path is generated at the classical level and the
various masses are proportional to the weak coupling $\ld$.

\subsection{Results}\label{num3}

\begin{figure}[!t]\vspace*{-0.45cm}
\includegraphics[height=3.18in,angle=-90]{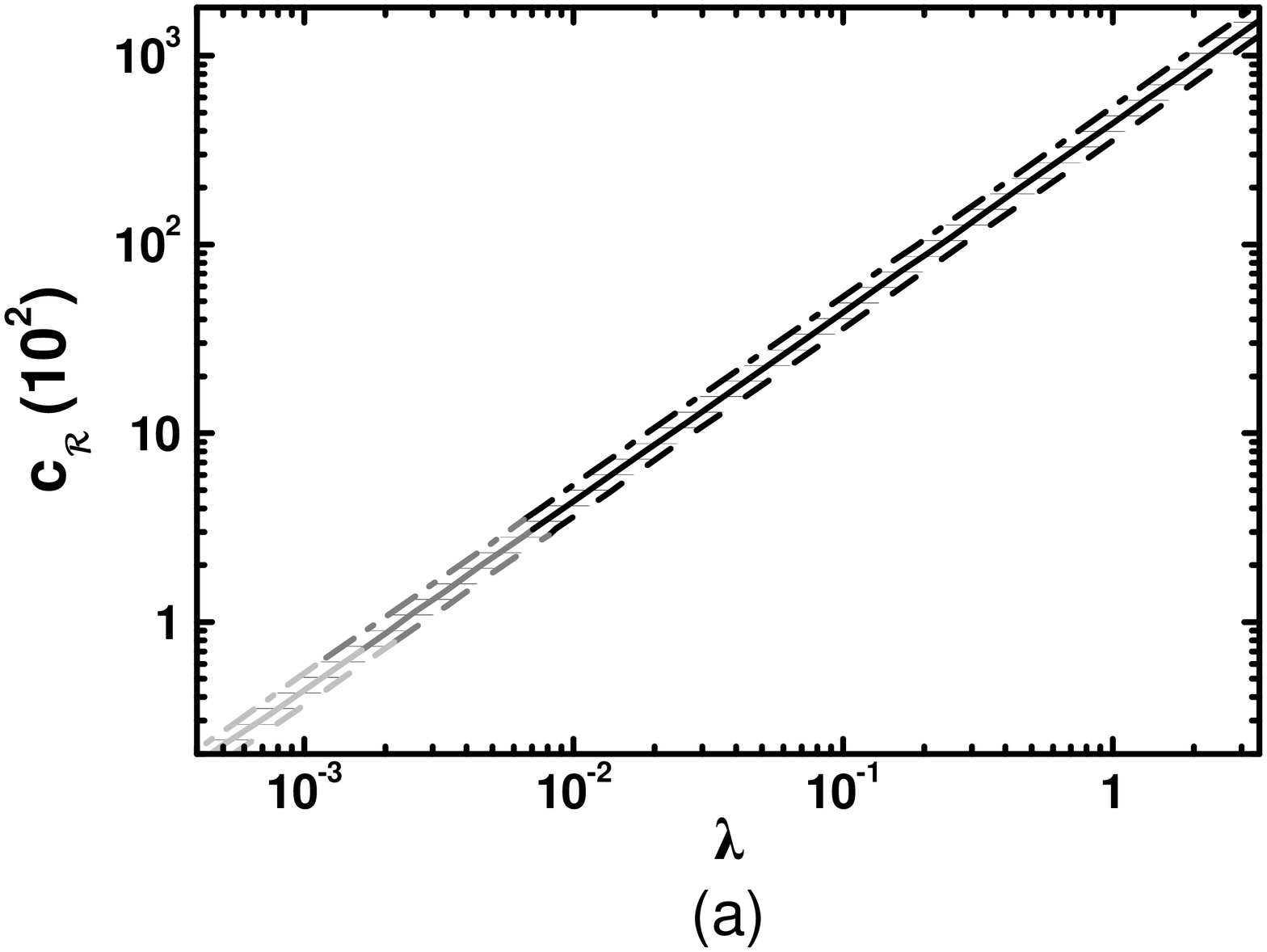}\hfil
\includegraphics[height=3.18in,angle=-90]{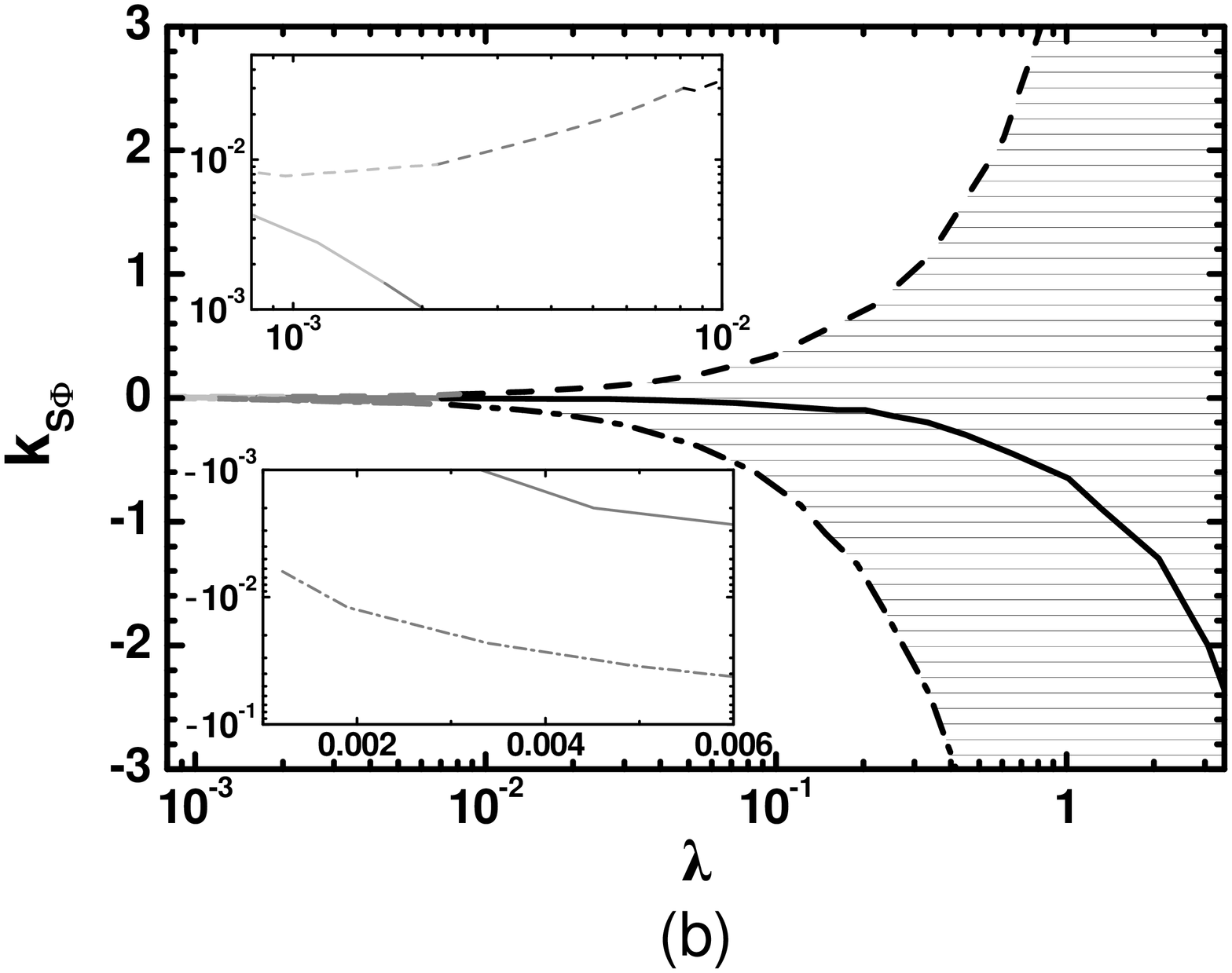}
\caption{\sl  Allowed (hatched) region as determined by
Eqs.~(1.5), (1.7) and (1.9) in the $\ld-\ck$ [$\ld-\ksp$] plane
(a) [(b)], for $\ks=\kpp=0.5$. The conventions adopted for the
type and the color of the various lines are described in Fig.
1.}\label{fig1}
\end{figure}


As can be easily seen from the relevant expressions above, the
model depends on the following parameters:
$\ld,\>\ks,~\kpp,~\ksp,\>\mbox{and}\>\ck$. Recall that we use
$\Trh=10^8~\GeV$ throughout. Our results are essentially
independent of $\ks$, provided that  $m_{\what S}^2>0$ for $\ld<1$
-- see in \Tref{tab2}. We therefore set $\ks=0.5$. Also we take a
central value $\kpp=0.5$. Besides these two values, in our
numerical code, we use as input parameters $\ck,~\ksp$ and
$\sg_*$. For every chosen $\ck\geq1$, we restrict $\ld$ and
$\sg_*$ so that the conditions \eqs{Nhi}{Prob} are satisfied. By
adjusting $\ksp$ we can achieve $n_{\rm s}$'s in the range of
Eq.~(\ref{nswmap}). Our results are displayed in \sFref{fig1}{a}
[\sFref{fig1}{b}], where we delineate the hatched region allowed
by \eqss{Nhi}{Prob}{nswmap} in the $\ld-\ck$ [$\ld-\ksp$] plane.
The conventions adopted for the various lines are shown in
\Fref{fig0}. In particular, the dashed [dot-dashed] lines
correspond to $n_{\rm s}=0.975$ [$n_{\rm s}=0.946$], whereas the
solid lines are obtained by fixing $n_{\rm s}=0.96$ -- see
Eq.~(\ref{nswmap}). The constraint of \sEref{Vl}{b} is satisfied
along the various curves whereas \sEref{Vl}{a} is valid only along
the gray and light gray segments of these. Along the light gray
segments, though, we obtain $\sg_*\geq\mP$. Note that for
vanishing $\ksp$ and $\kpp$ our results can be approximated by the
analytical expressions exhibited in the paragraph (ii) of
\Sref{nmi1} replacing $\sqrt{\ld}$ with $\ld$. Indeed, $\ck$
remains almost proportional to $\ld$ and for constant $\ld$, $\ck$
increases as $\ns$ decreases. We remark that mostly negative
$\ksp$'s are needed which for
$\ld>0.16\Leftrightarrow\sg_*<0.1\mP$ -- see \Eref{sigs1} -- take
quite natural (of order one) values. Focusing on $\ksp<0$ for
$n_{\rm s}=0.96$ and $\Ne_*\simeq50$ we find
\beq\label{res1} 112\lesssim
\ck\lesssim1.6\cdot10^5\>\>\>\mbox{with}\>\>\>2.5\cdot10^{-3}\lesssim
\ld\lesssim3.7\>\>\>\mbox{and}\>\>\> 0\lesssim
-\ksp\lesssim2.5\,.\eeq
Also $6.8\lesssim {|\as|/10^{-4}}\lesssim8.2$ and $r\simeq3.8\cdot
10^{-3}$ which lie within the allowed ranges of \Eref{nswmap}.

\section{nMI with Quartic Potential for a Gauge non-Singlet Inflaton}\label{fhi1}

\begin{floatingtable}[r]
\begin{tabular}{|l||lll|}\hline
{\sc Superfields}&$S$&$\Phi$&$\bar\Phi$\\\hline\hline
$U(1)_{B-L}$&$0$&$1$&$-1$\\\hline
$R$ &$1$&$0$&$0$\\\hline
\end{tabular}
\caption {\sl Charge assignments of the superfields.}\label{ch}
\end{floatingtable}
In the present scheme, the inflaton field can be identified with
the radial component of a conjugate pair of Higgs superfields. We
here -- cf. \cref{jones2} -- focus on the Higgs superfields,
$\bar\Phi$ and $\Phi$, with $B-L=-1,~1$ which break the GUT
symmetry $\Ggut=\Gsm\times U(1)_{B-L}$ down to MSSM gauge group
$\Gsm$ through their VEVs. We also impose a $U(1)$ R symmetry,
$U(1)_R$, which guarantees the linearity of the superpotential,
$W$, w.r.t the singlet $S$ -- see \Tref{ch}. The functions $W$,
$\Fcr$ and $\Fk$, which respect the imposed symmetries, are
respectively given by
 \beqs \bea \hspace*{-2.2cm} W&=&\ld S\lf\bar\Phi\Phi-\Mpq^2\rg,\>\>\>\Fcr={c_{\bar\Phi\Phi}\over2\mP^2}\bar\Phi\Phi,\label{W2}\\
\Fk&=&{|S|^2\over\mP^2}+{|\Phi|^2\over\mP^2}+{|\bar\Phi|^2\over\mP^2}-\ks{|S|^4\over\mP^4}-
\kpp{|\Phi|^4\over\mP^4}-\kppb{|\bar\Phi|^4\over\mP^4}\nonumber\\&-&2\ksp
{|S|^2|\Phi|^2\over\mP^4}-2\kspb{|S|^2|\bar\Phi|^2\over\mP^4}-2\kppp{|\Phi|^2|\bar\Phi|^2\over\mP^4}\,\cdot\label{W1}
\eea \eeqs
We below outline the salient features of our inflationary scenario
(\Sref{fhi11}) and then, we present its predictions in
Sec.~\ref{num1}.

\subsection{Structure of the Inflationary Potential}\label{fhi11}

The EF F--term (tree level) SUGRA scalar potential, $\Vhio$, for
this model is obtained by substituting \eqs{W2}{W1} into
\eqss{Vsugra}{Omg1}{Omg} with $\Phi^1=S,~\Phi^2=\Phi$ and
$\Phi^3=\bar\Phi$. If we parameterize the SM neutral fields $\Phi$
and $\bar\Phi$ by
\beq\label{hpar} \Phi=\sg
e^{i\th}\cos\theta_\Phi/\sqrt{2}\>\>\>\mbox{and}\>\>\>\bar\Phi=\sg
e^{i\thb}\sin\theta_\Phi/\sqrt{2},\eeq
we can easily deduce that a D-flat direction occurs at
\beq\label{inftr}
\th=\thb=0\>\>\>\mbox{and}\>\>\>\thn={\pi/4},\eeq
provided that $K$ remains invariant under the interchange $\Phi
\to \bar\Phi\>\>\>\mbox{and}\>\>\>\bar\Phi\to\Phi,$ which implies
\beq\kpp=\kppb\>\>\>\mbox{and}\>\>\>\ksp=\kspb.\label{conj2}\eeq
From \Eref{Vsugra}, we can verify that for $\ck\gg1$, $\ksp\ll1$
and $M\ll\mP$, $\Ve$ takes a form suitable for the realization of
nMI, since it develops a plateau. The (almost) constant potential
energy density $\Vhio$ and the corresponding Hubble parameter
$\He_{\rm I}$ -- along the trajectory in \Eref{inftr} -- are given
by
\beqs\beq \Vhio=\mP^4\frac{\ld^2(\xsg^2-4\mpq^2)^2}{16\fsp
f_R^2}\simeq{\ld^2\mP^4\over16\fsp\ck^2}~~~\mbox{and}~~~ \He_{\rm
I}={\Vhio^{1/2}\over\sqrt{3\fsp}\mP}\simeq{\ld\mP\over4\sqrt{3\fsp}\ck}\,,
\label{1Vhio}\eeq
where $\fr$ and $\fsp$, given by their definitions below
\Eref{Vhig}, are specified as follows
\beq \label{Vhi1} \fr=1+\ck
\xsg^2+(\kpp+\kppp){\xsg^4\over24},~\fsp=1-\ksp\xsg^2~~~
\mbox{with}~~~\mpq=\frac{\Mpq}{\mP}~~
\mbox{and}~~\ck=\frac{c_{\bar\Phi\Phi}}{4}-\frac{1}{6}\,\cdot\eeq\eeqs

\renewcommand{\arraystretch}{1.4}
\begin{table}[!t]
\begin{center}
\begin{tabular}{|c||c|l|}\hline
{\sc Fields}&{\sc Eigenstates}&\hspace*{2.cm} {\sc Masses Squared}\\
\hline\hline
2 real scalars&$\widehat \theta_\Phi$ &$ m_{\widehat
\theta_\Phi}^2\simeq g^2\mP^2\xsg^2/\fp\fr$\\
&$\widehat\theta_{+}$&$m_{\widehat\theta+}^2=\ld^2\mP^2\xsg^6\ck/12\fr^3\fsp\simeq4\Hhi^2$\\
1 complex scalar&$\widehat S$ &$ m_{\widehat s}^2=\ld^2\mP^2\lf
12 + \xsg^2 \lf1+6\ck^2\xsg^2\rg\lf6 \ks -1\rg\right.$\\
&&$\left.+36\ck^3\ks\xsg^6\rg/144\ck^2\fsp^3\fr^2$\\\hline
1 gauge  boson& $A_{BL}$ &
$ M_{BL}^2=g^2\mP^2\xsg^2\fp /\fr$\\
\hline
$4$ Weyl spinors & $\what \psi_\pm={\what{\psi}_{\Phi+}\pm
\what{\psi}_{S}\over\sqrt{2}}$ & $m^2_{\what
\psi\pm}\simeq\ld^2\mP^2\lf2+\ksp\xsg^2(\ck\xsg^2-1)\rg/36\fsp^3\fr^2\ck^2$\\
&$\ldu_{BL}, \widehat\psi_{\Phi-}$&$M_{BL}^2=g^2\mP^2\xsg^2\fp /\fr$\\
\hline
\end{tabular}\end{center}
\caption{\sl\ftn The mass spectrum of the model along the
inflationary trajectory of Eq.~(4.3) in the presence of the
conditions in Eq.~(4.4). To avoid very lengthy formulas we neglect
terms proportional to $\mpq^2\ll\xsg^2$. }\label{tab1}
\end{table}

We next proceed to check the stability of the trajectory in
\Eref{inftr} w.r.t the fluctuations of $\thn,~\th,~\thb$ and
$S=s/\sqrt{2}$ -- here $S$ has been rotated to the real axis via a
suitable R transformation. We first check that $\Ve_{,\chi}=0$,
with $\chi=\thn,\th,\thb$ and $s$, and then we find
\bea \nonumber K_{\al\bbet}\dot\Phi^\al \dot\Phi^{*\bbet}&=&
{1\over2}J^2\lf\dot \sg ^2+{1\over2}\sg^2\dot\theta^2_+
\rg+{\fp\sg^2\over 2\fr}\lf{1\over2}\dot\theta^2_-
+\dot\theta^2_\Phi \rg+\frac{\fsp}{2\fr}\dot s\\
&=&\frac12\lf\dot{\widehat \sg}^2+\dot{\widehat
\th}_+^2+\dot{\widehat \th}_-^2+\dot{\widehat
\th}_\Phi^2+\dot{\widehat s}^2\rg\>\>\>\mbox{with}\>\>\>
\th_{\pm}=\frac{\bar\th\pm\th}{\sqrt{2}}\>\>\>\mbox{and}\>\>\>s={S\over\sqrt{2}}\,\cdot\label{Snik}\eea
In the last line, we introduce the EF canonically normalized
fields, $\se, \widehat \theta_+,\widehat\theta_-,
\widehat\theta_\Phi$ and $\widehat s$, defined as follows -- cf.
\cref{nmH}:
\beq \label{cannor}
\frac{d\se}{d\sg}=J\simeq{\sqrt{6}\over\xsg},\>\>\widehat \theta_+
={J\sg\theta_+\over\sqrt{2}},\>\>\widehat \theta_-
={\sqrt{\fp\over2\fr}}\sg\theta_-,\>\>\widehat \theta_\Phi =
\sqrt{\frac{\fp}{\fr}}\sg\lf\theta_\Phi-{\pi\over4}\rg~~\mbox{and}~~\widehat
s=\sqrt{\frac{\fsp}{\fr}}s, \eeq
where $J$ and $\fk$ can be found from \eqs{VJe}{fkdef}. Also
$\fp=1-\kpp\xsg^2$.

Having defined the canonically normalized scalar fields, we can
derive the mass spectrum of the model along the direction of
\Eref{inftr}. Our results are listed in \Tref{tab1}, where we
present the eigenvalues and the corresponding eigenvectors of the
relevant mass-squared matrices. As we observe, no instability
arises in the spectrum, since $\ks\gtrsim1$ ensures $m_{\widehat
S}^2>0$ and $m^2_{\what \theta_\Phi}>0$ thanks the D-term
contributions which are proportional to $g\simeq0.7>\ld$.
Moreover, the masses of the various scalars remain greater than
$\Hhi$ and so any perturbations of the fields other than the
inflaton are safely eliminated. We also remark that $U(1)_{B-L}$
is broken during nMI and therefore the gauge boson $A_{BL}$
becomes massive absorbing the massless Goldstone boson associated
with $\what\th_-$. As a consequence, no cosmic strings are
produced at the end of nMI and so, no extra restrictions on the
parameters have to be imposed. From \Tref{tab1} we can deduce that
the numbers of bosonic (8) and fermionic (8) d.o.fs are equal.
Plugging these results into \Eref{Vhic} we get $\Delta V$, which
is dominated by the contributions from $m_{\widehat \theta_\Phi}$
and $M_{BL}$ since these are proportional to $g\gg\ld$. Namely we
find
\beq  \Delta V\simeq{1\over64\pi^2}\lf m_{\widehat
\theta_\Phi}^4\ln{(m_{\widehat \theta_\Phi}^2/\Lambda^2)}
-M_{BL}^4\ln{(M_{BL}^2/\Lambda^2)}\rg. \label{Vrc} \eeq
Note that the presence of $\fp\neq1$ prevents the exact
cancellation, occurring in \cref{nmH}, of the two contributions
above. Since this result can be continued until the SUSY vacuum of
the theory, we determine there the employed  $\Lambda$ by imposing
the condition $\Delta V=0$.

\subsection{Results}\label{num1}

The free parameters of this model are
$\ld,~M,\>\ks,~\kpp,~\kppp,~\ksp,\>\mbox{and}\>\ck.$ Following the
same reasoning with \Sref{num3} we set $\ks=1$ and $\kpp=0.1$
throughout. Moreover, we can determine $\Mpq$ identifying the mass
of the gauge boson $A_{BL}$ in the SUSY vacuum with the GUT scale
$\Mgut\simeq2\cdot10^{16}~\GeV$ within the MSSM, i.e.
\beq \label{unif} \sqrt{{f_{\Phi0}/
f_{R0}}}g\Mpq=\Mgut\>\Rightarrow\>\mpq\simeq(2\sqrt{2c^{\rm
max}_{R}-\ck})^{-1}~~\mbox{with}~~ c^{\rm max}_{R}={g^2\mP^2/
8\Mgut^2},\eeq
$f_{R0}=f_{R}(\xsg=2\mpq)$ and $f_{\Phi0}=f_{\Phi}(\xsg=2\mpq)$.
The requirement $2c^{\rm max}_{R}-\ck>0$ sets an upper bound
$\ck<2c^{\rm max}_{R}\simeq1.8\cdot10^3$. To obtain solutions
consistent with the requirements of \Sref{obs} we need a rather
low $\kpp$. We take $\kpp=0.01$.

\begin{figure}[!t]\vspace*{-0.45cm}
\includegraphics[height=3.18in,angle=-90]{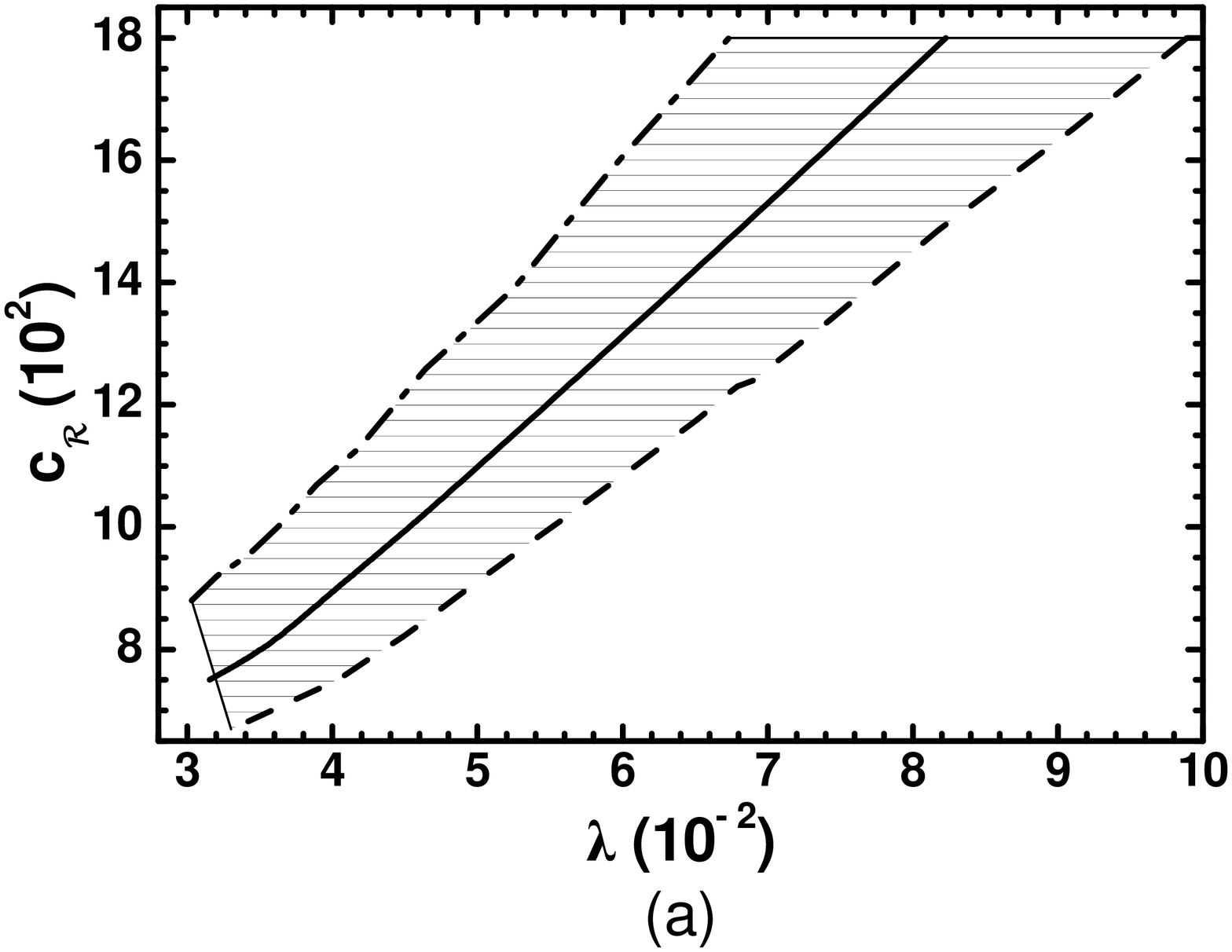}\hfil
\includegraphics[height=3.18in,angle=-90]{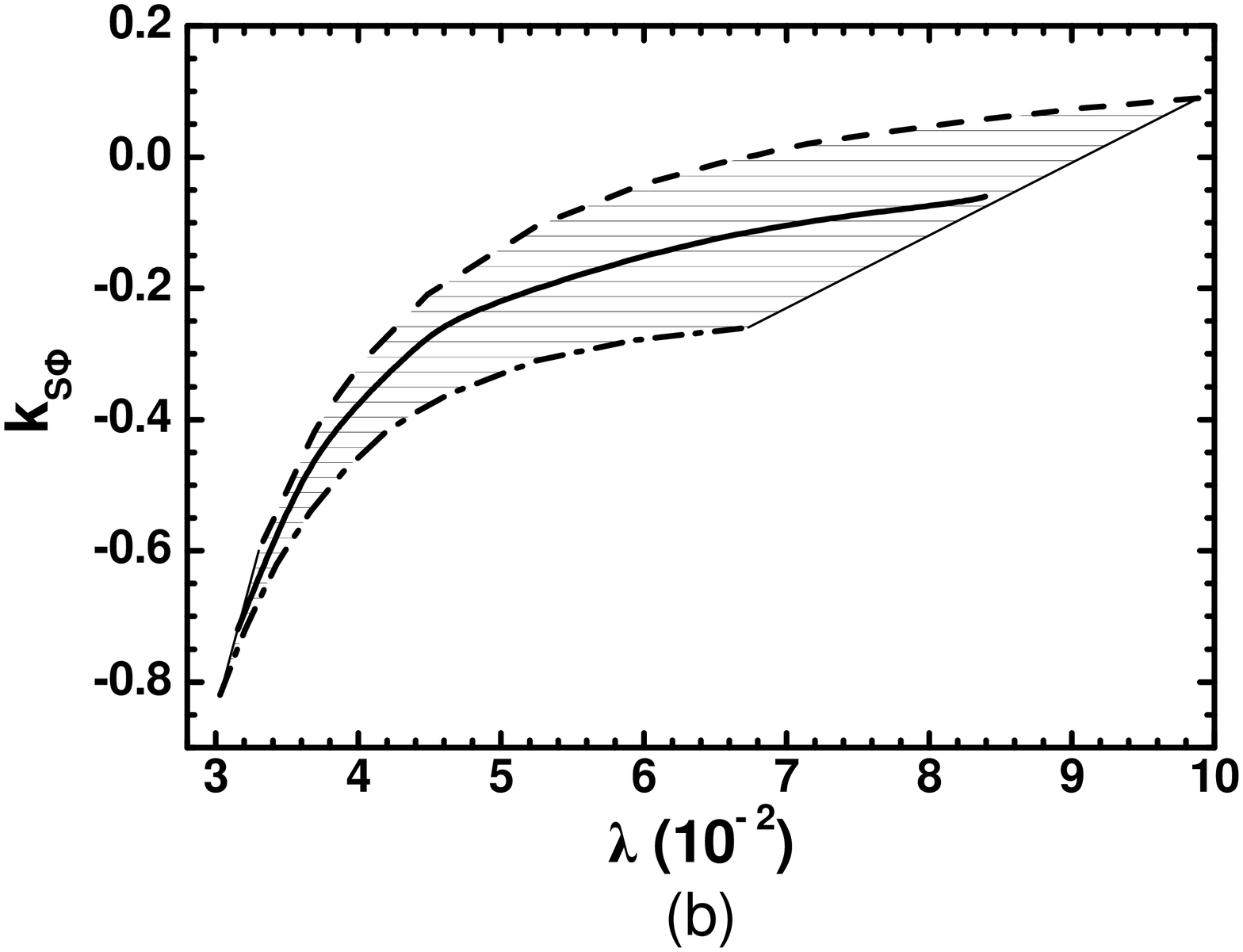}
\caption{\sl  Allowed (hatched) region as determined by
Eqs.~(1.5), (1.7) and (1.9) in the $\ld-\ck$ [$\ld-\ksp$] plane
(a) [(b)] for $\ks=1$, $\kppp=0.1$ and $\kpp=0.01$. The
conventions adopted are described in Fig. 1.}\label{fig2}
\end{figure}


Working along the lines of \Sref{num3} we delineate the allowed
(hatched) region, as determined  by \eqss{Nhi}{Prob}{nswmap} in
the $\ld-\ck$ [$\ld-\ksp$] plane -- see \sFref{fig2}{a}
[\sFref{fig2}{b}]. The conventions adopted for the various lines
are shown in \Fref{fig0}. We observe that the allowed region is
considerably shrunk w.r.t that obtained with a gauge singlet
inflaton in \Fref{fig1} and this is limited in the regime where
only \sEref{Vl}{b} is valid. Also $\ck$ remains almost
proportional to $\ld$ and increases as $\ns$ decreases. Note that
for vanishing $k$'s, our results can be approximated by the
analytical expressions exhibited in the paragraph (ii) of
\Sref{nmi1} replacing $\sqrt{\ld}$ with $\ld/2$. Especially for
$n_{\rm s}=0.96$ and $\Ne_*\simeq50$ we find
\beq\label{res2} 7.3\lesssim
\ck/10^2\lesssim18\>\>\>\mbox{with}\>\>\>2.9\lesssim
\ld/10^{-2}\lesssim8.4\>\>\>\mbox{and}\>\>\> 0.6\lesssim
-\ksp/10^{-1}\lesssim8.5.\eeq
Therefore, $\ksp<0$ assists us again to obtain the central $\ns$
in \sEref{nswmap}{a}. Also $0.7\lesssim
{|\as|/10^{-3}}\lesssim1.2$ and $r\simeq(3-3.8)\cdot 10^{-3}$ in
agreement with \Eref{nswmap}.

\section{nMI with Quadratic Potential}\label{fhi2}

This model \cite{nMCI} is characterized by the following
ingredients:
\beq \label{mfhi2} W=m S\Phi,~\Fcr={c_\Phi\over \sqrt{2}\mP}\Phi,~
\Fk={|S|^2\over\mP^2}+{|\Phi|^2\over\mP^2}-2\ks{|S|^4\over\mP^4}-
2\kpp{|\Phi|^4\over\mP^4}-2\ksp
{|S|^2|\Phi|^2\over\mP^4}\,\cdot\eeq
To ensure the form of $W$ and $\Fk$ we can impose a global $U(1)$
symmetry under which $S$ and $\Phi$ have charges $1$ and $-1$. The
imposed $U(1)$ is broken though during nMI by $\Fcr$. Due to the
form of $\Fcr$, in which $\Phi$ appears linearly, $\Phi$ has to be
singlet under the gauge group of the theory and therefore, we
obtain D$^\al=0$ by construction. In \Sref{fhi21} we outline the
derivation of the inflationary potential of this model and in
\Sref{num2} we present our results.

\subsection{Structure of the Inflationary
Potential}\label{fhi21}

Inserting \Eref{mfhi2} into \eqss{Omg}{Omg1}{Vsugra} we can derive
$\Ve$ which, along the trajectory of \Eref{inftr3} for $\ck\gg1$
and $\xsg\ll\sqrt{6}$, develops a plateau with (almost) constant
potential energy density, $\Vhio$ and corresponding Hubble
parameter $\He_{\rm I}$ calculated, in accordance with
\Eref{Vhig}, as
\beqs\beq\label{2Vhio} \Vhio=
{m^2\sg^2\over2\fsp\fr^2}\simeq{m^2\mP^2\over2\ck^2\fsp}\>\>\>\mbox{and}\>\>\>
\He_{\rm
I}={\Vhio^{1/2}\over\sqrt{3}\mP}\simeq{m\over\sqrt{6\fsp}\ck},\eeq
where $\fr$ and $\fsp$, defined below \Eref{Vhig}, are specified
as follows
\bea
\fr=1+\ck\xsg-{\xsg^2/6}+\kpp\xsg^4/6\>\>\>\mbox{and}\>\>\>\fsp=1-\ksp\xsg^2\,.
\eea\eeqs
To check the stability of the trajectory of \Eref{inftr3} w.r.t
the fluctuations of the various fields we expand $\Phi$ and $S$ in
real and imaginary parts according to the prescription of
\Eref{cannor3a}. We then find that along the trajectory of
\Eref{inftr3}, $K_{\al\bbet}$ defined in \Eref{Kab} turns out to
be
\bea \lf K_{\al\bbet}\rg=\diag\lf
J^2,\fsp/\fr\rg\>\>\>\mbox{where}\>\>\> J\simeq
{\sqrt{3}\ck/\sqrt{2} \fr}\simeq
\sqrt{3/2}{\xsg^{-1}}\,.\label{VJe2}\eea
Consequently, we can introduce the EF canonically normalized
fields, $\se, \widehat \th$ and $\widehat s_i$, with $i=1,2$ which
satisfy \Eref{K3} with the hatted fields defined as in
\Eref{cannor3b}. As in the previous cases, we check the stability
of the inflationary path in \Eref{inftr3} deriving the mass
spectrum of the model, presented in \Tref{tab2}. We observe that
now the various masses are proportional to $m$ and obviously our
findings are similar to those obtained in \Sref{fhi31}.
%

\renewcommand{\arraystretch}{1.5}

\begin{table}[!t]
\bec\begin{tabular}{|c|c|c|}\hline
{\sc Fields} &{\sc Eingestates} & {\sc Masses Squared}\\ \hline
\hline
$1$ real scalar &$\what \th$ & $m_{\what \th}^2\simeq\ck m^2\xsg/\fr^3J^2\simeq4\Hhi^2$\\
$2$ real scalars &$\what s_1, \what s_2$ & $m_{\what s}^2\simeq
m^2\lf 2+\ck^2\xsg^2(12\ck\ks-1)\rg/\fsp^3\fr^2(2+3\ck^2)$\\\hline
$2$ Weyl spinors & $\what \psi_\pm={\what{\psi}_{\Phi}\pm
\what{\psi}_{S}\over\sqrt{2}}$& $m_{\what \psi\pm}^2\simeq
m^2(6+\xsg^2+6\ck\ksp\xsg^3)^2/12\fsp^3\fr^2(2+3\ck^2)$\\
\hline
\end{tabular}\eec
\hfill \caption{\sl The mass spectrum of the model along the
inflationary trajectory in Eq. (3.2).}\label{tab2}
\end{table}

\subsection{Results}\label{num2}

\begin{figure}[!t]\vspace*{-0.45cm}
\includegraphics[height=3.18in,angle=-90]{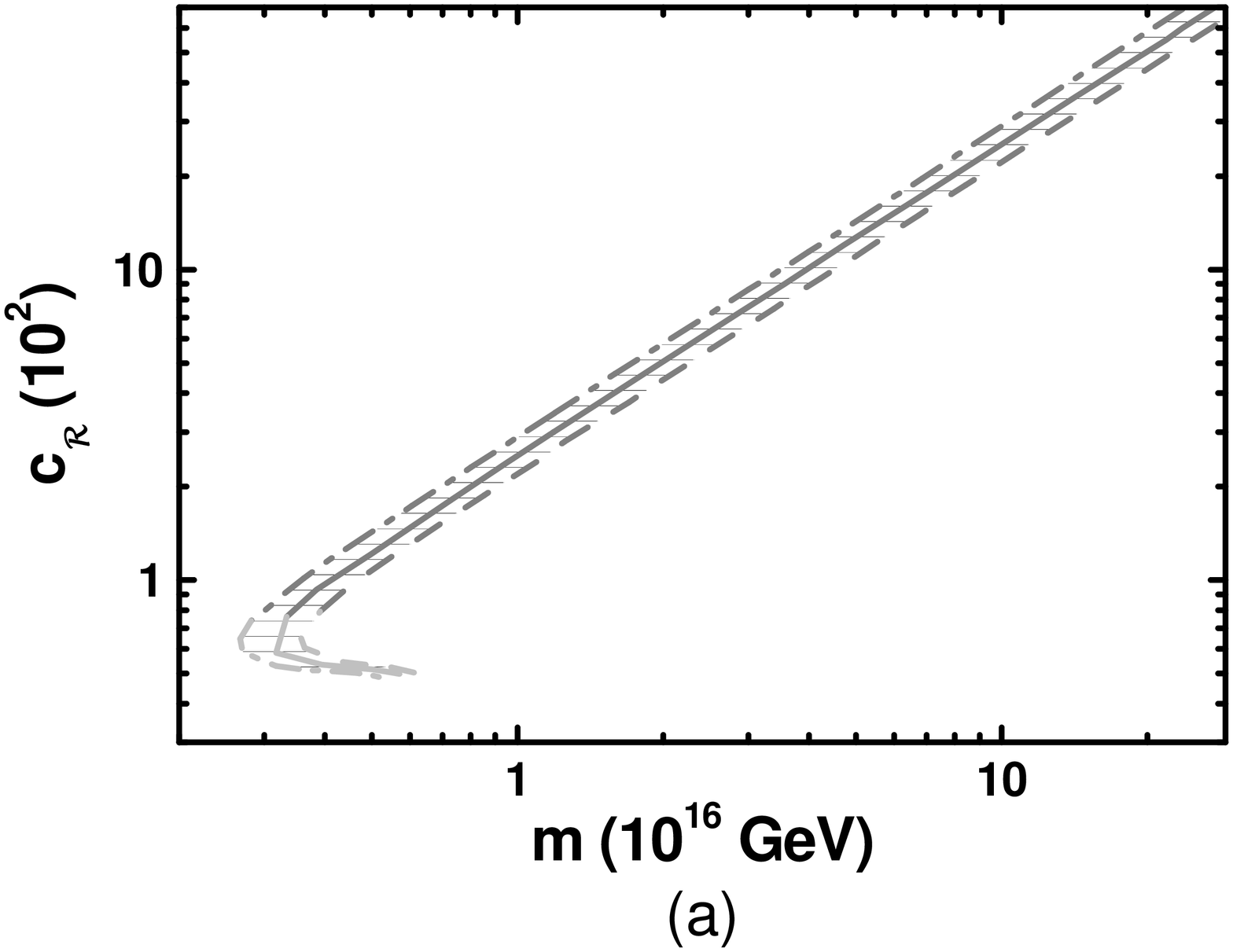}\hfil
\includegraphics[height=3.18in,angle=-90]{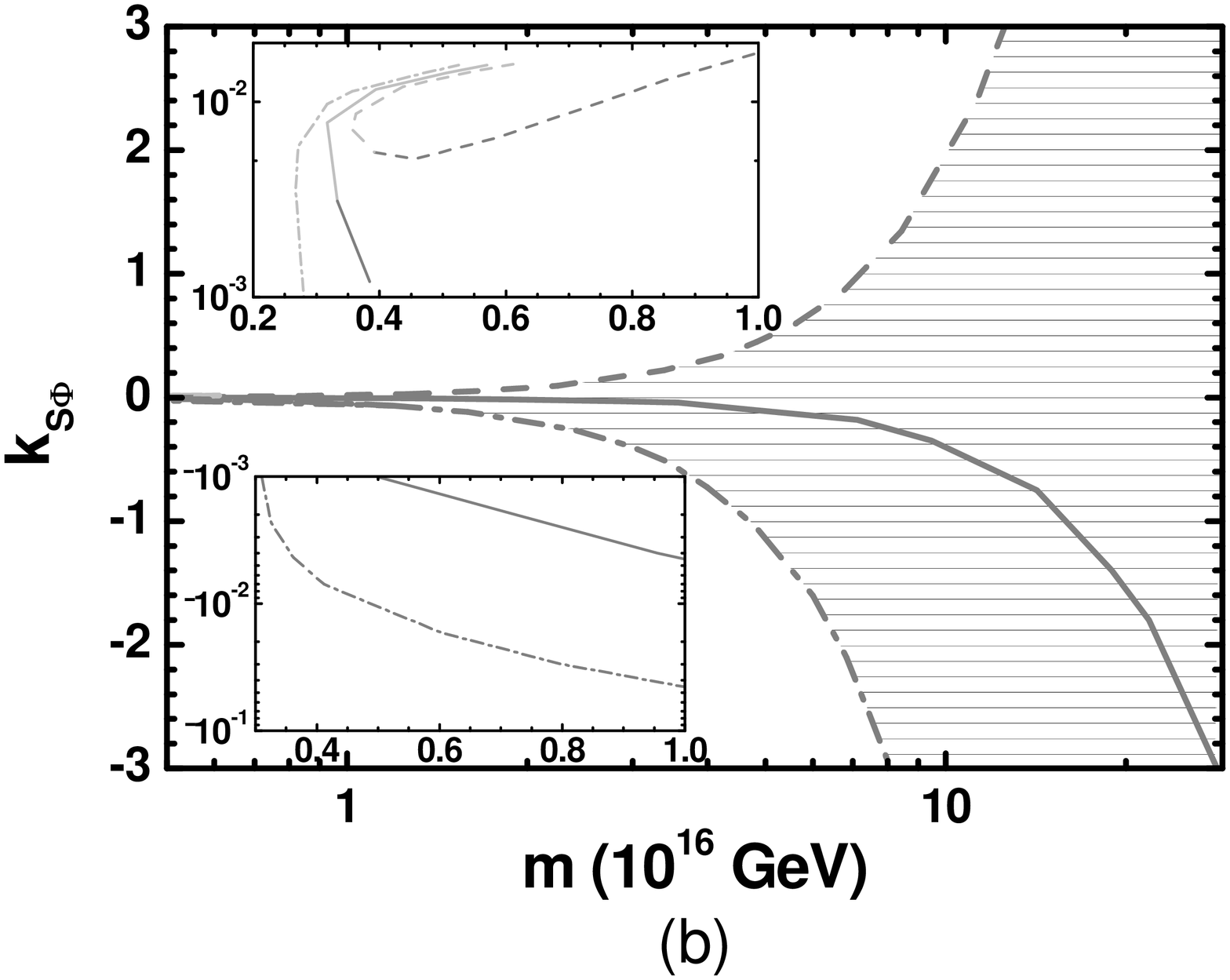}
\caption{\sl  Allowed (hatched) region as determined by
Eqs.~(1.5), (1.7) and (1.9) in the $m-\ck$ [$m-\ksp$] plane (a)
[(b)] for $\ks=\kpp=0.5$. We follow the conventions of
Fig.~1.}\label{fig3}
\end{figure}


This model depends on the parameters:
$m,\>\ks,~\ksp,~\kpp,\>\mbox{and}\>\ck.$ Following the same
reasoning with \Sref{num3} we set throughout $\ks=\kpp=0.5$. The
allowed (hatched) regions in the $m-\ck$ and $m-\ksp$ plane, as
determined by \eqss{Nhi}{Prob}{nswmap}, are depicted in
\sFref{fig3}{a} and \sFref{fig3}{b} respectively, following the
conventions of \Fref{fig0}. We confine conservatively ourselves to
values $m\lesssim3\cdot10^{17}~\GeV$. For $\ksp\simeq0$ and
$\kpp\simeq0$, our results here are almost identical with those in
paragraph (ii) of \Sref{nmi2}. For $m\gtrsim4\cdot10^{16}~\GeV$,
rather natural $\ksp$'s result to $n_{\rm s}=0.96$. For this
$\ns$, $\Ne_*\simeq50$ and confining ourselves to $\ksp<0$ we find
\beq\label{res3} 1.2\lesssim
\ck/10^2\lesssim60\>\>\>\mbox{with}\>\>\>0.5\lesssim
m/10^{16}~\GeV\lesssim3\>\>\>\mbox{and}\>\>\> 0\lesssim
-\ksp\lesssim2.4.\eeq
Also ${\as}\simeq-6.5\cdot10^{-4}$ and $r\simeq3.9\cdot10^{-3}$
compatibly with \Eref{nswmap}. Obviously, the most important
advantage of the present model is that it is consistent with the
criterion of \sEref{Vl}{a}.

\section{Conclusions}\label{con}

We reviewed the implementation of nMI in both a non-SUSY and a
SUSY framework, by conveniently choosing the coupling function of
the inflaton to gravity. In all cases the inflationary observables
are compatible with the current data. In the SUSY cases the
tachyonic instability, occurring along the direction of the
accompanying non-inflaton field, can be remedied by considering
terms up to the fourth order in the kinetic part of the frame
function. Some of these terms assist in obtaining $\ns$ close to
its central observationally favored value. The role of the
inflaton can be played by a gauge singlet or non-singlet
superfield. In the latter case, though, the flatness of the
potential is very sensitive to the higher order terms and the
radiative corrections. For this reason, we conclude that nMI
within SUGRA is more naturally realized by a gauge singlet
inflaton, especially when it couples linearly to gravity, since
then the effective theory is valid up to $\mP$.


\def\ijmp#1#2#3{{\emph{Int. Jour. Mod. Phys.}}
{\bf #1},~#3~(#2)}
\def\plb#1#2#3{{\emph{Phys. Lett.  B }}{\bf #1},~#3~(#2)}
\def\zpc#1#2#3{{Z. Phys. C }{\bf #1},~#3~(#2)}
\def\prl#1#2#3{{\emph{Phys. Rev. Lett.} }
{\bf #1},~#3~(#2)}
\def\rmp#1#2#3{{Rev. Mod. Phys.}
{\bf #1},~#3~(#2)}
\def\prep#1#2#3{\emph{Phys. Rep. }{\bf #1},~#3~(#2)}
\def\prd#1#2#3{{\emph{Phys. Rev.  D} }{\bf #1},~#3~(#2)}
\def\npb#1#2#3{{\emph{Nucl. Phys.} }{\bf B#1},~#3~(#2)}
\def\npps#1#2#3{{Nucl. Phys. B (Proc. Sup.)}
{\bf #1},~#3~(#2)}
\def\mpl#1#2#3{{Mod. Phys. Lett.}
{\bf #1},~#3~(#2)}
\def\arnps#1#2#3{{Annu. Rev. Nucl. Part. Sci.}
{\bf #1},~#3~(#2)}
\def\sjnp#1#2#3{{Sov. J. Nucl. Phys.}
{\bf #1},~#3~(#2)}
\def\jetp#1#2#3{{JETP Lett. }{\bf #1},~#3~(#2)}
\def\app#1#2#3{{Acta Phys. Polon.}
{\bf #1},~#3~(#2)}
\def\rnc#1#2#3{{Riv. Nuovo Cim.}
{\bf #1},~#3~(#2)}
\def\ap#1#2#3{{Ann. Phys. }{\bf #1},~#3~(#2)}
\def\ptp#1#2#3{{Prog. Theor. Phys.}
{\bf #1},~#3~(#2)}
\def\apjl#1#2#3{{Astrophys. J. Lett.}
{\bf #1},~#3~(#2)}
\def\n#1#2#3{{Nature }{\bf #1},~#3~(#2)}
\def\apj#1#2#3{{Astrophys. J.}
{\bf #1},~#3~(#2)}
\def\anj#1#2#3{{Astron. J. }{\bf #1},~#3~(#2)}
\def\mnras#1#2#3{{MNRAS }{\bf #1},~#3~(#2)}
\def\grg#1#2#3{{Gen. Rel. Grav.}
{\bf #1},~#3~(#2)}
\def\s#1#2#3{{Science }{\bf #1},~#3~(#2)}
\def\baas#1#2#3{{Bull. Am. Astron. Soc.}
{\bf #1},~#3~(#2)}
\def\ibid#1#2#3{{\it ibid. }{\bf #1},~#3~(#2)}
\def\cpc#1#2#3{{Comput. Phys. Commun.}
{\bf #1},~#3~(#2)}
\def\astp#1#2#3{{Astropart. Phys.}
{\bf #1},~#3~(#2)}
\def\epjc#1#2#3{{Eur. Phys. J. C}
{\bf #1},~#3~(#2)}
\def\nima#1#2#3{{Nucl. Instrum. Meth. A}
{\bf #1},~#3~(#2)}
\def\jhep#1#2#3{{\emph{JHEP} }
{\bf #1},~#3~(#2)}
\def\jcap#1#2#3{{\emph{JCAP} }
{\bf #1},~#3~(#2)}

\end{document}